\documentclass[pre,twocolumn,showpacs,preprintnumbers,amsmath,amssymb,floatfix]{revtex4}
\usepackage[dvips]{graphicx}    % For graphics
\usepackage{ulem}       % For double underlining
\usepackage{natbib}

\usepackage{color} %to highlight changes
\newcommand{\be}{\begin{equation}}      % For displaying equations
\newcommand{\bea}{\begin{eqnarray}}
\newcommand{\ee}{\end{equation}}
\newcommand{\eea}{\end{eqnarray}}
\newcommand{\bdm}{\begin{displaymath}}
\newcommand{\edm}{\end{displaymath}}
\newcommand{\half}{{\textstyle \frac{1}{2}}}    % For fractions

\newcommand{\ten}[1]{\uuline{#1}{}}     % For tensors

\renewcommand{\vec}[1]{{\mathbf #1}}        % For vectors
\renewcommand{\det}[1]{{\rm Det}[#1]}% For determinants

\def\sp#1{^{\rm #1} }

\begin{document}

%%%%%%%%%%%%%%%%  TITLE AND AUTHOR DETAILS  %%%%%%%%%%%%%%%%%%%
\title{On the physical basis for the nematic rubber elastic free energy}
\author{J. S. Biggins M. Warner}
\affiliation{Cavendish Laboratory, University of Cambridge,
Madingley Road, Cambridge CB3 0HE, U.K.} \date{\today}

%%%%%%%%%%%%%%%%%%%%%%%%   ABSTACT   %%%%%%%%%%%%%%%%%%%%%%%%%%
\begin{abstract}
We discuss why it is physical to keep terms in the nematic rubber elastic free energy that reflect
the order parameter dependence of the natural size of the network polymers.  We address a point of
difficulty in some mathematical approaches to this problem.
\end{abstract}
 \maketitle

%%%%%%%%%%%%%%%%%%%%%   MAIN BODY   %%%%%%%%%%%%%%%%%%%%%%%%%%%

\section{Introduction}

Nematic elastomers are formed by crosslinking nematic polymers.  Such molecules order orientationally
because of rod-like elements incorporated in their main chains (MC polymers) or pendant as side
chains (SC).  In both cases orientational order induces the backbones to change their mean shapes
from spherical to elongated (prolate) or flattened (oblate) forms.  Since chain shape and the
macroscopic shape of a network are intimately related, there is a strong coupling between the nematic
order and the shape of such solids.  This obtains for both prolate and oblate chain shapes -- our
conclusions will hold for both types of polymer.  Several elastic phenomena, unique to nematic
elastomers, arise from this coupling.  The most relevant in a discussion of  changing the magnitude
of the order $Q$, is that such changes induce huge elongations and contractions (of many 100s\%). The
order can be changed by changing temperature, or by illumination if at least some of the rods contain
chromophores.  The effect is reversible and is described in a monograph \cite{warnerbook:03} that
summarises the work of many research groups.  The magnitude of the spontaneous elongation,
$\lambda_m$, is proposed (as explained below) as an indicator of the shape anisotropy of the
polymers, $\lambda_m = (\ell_{\parallel}/\ell_{\perp})^{1/3}$, where $\ell_{\parallel}$ and
$\ell_{\perp}$ are the Flory effective step lengths parallel and perpendicular to the nematic
ordering direction, $\vec{n}$.

Our purpose here is to clarify the way in which, at constant temperature, the nematic can perhaps
change the magnitude as well as the direction of its order as distortions are imposed.  For instance,
close to the nematic-isotropic transition, or in regions of extreme distortion close to the core of a
disclination, $Q$ may reduce.  Some investigations, particularly from the mathematical community,
have had understandable concerns that the standard nematic rubber elastic free energy apparently has
terms unbounded from below in the limit $Q \rightarrow 1$.  To retain a physical description, authors
 have instead rigidly constrained such terms to be constant.  We point
out that (i) realistic chain models do have the critical terms varying, and (ii) that there are more
practical and physically-founded ways to deal with this apparent difficulty.

A simple extension\cite{warnerbook:03} to nematic elastomers of the classical Gaussian theory of
rubber elasticity yields a model, ideal free energy density:
 \be f_{\rm el} =\half \mu \left[ {\rm Tr} \left[
  \ten{\ell}_0\cdot \ten{\lambda}\sp{T}\cdot \ten{\ell}_\vec{n}^{-1}\cdot \ten{\lambda}\right]
 + \ln\left(\frac{\det{\ten{\ell}_0}}{\det{\ten{\ell}}}\right) \right]
\label{eqn:single} \ee
 where $\mu$ is the shear modulus of the rubber in the isotropic state.   $\ten{\ell}_0$ and
 $\ten{\ell}_\vec{n}$ are the Flory effective step length tensors that give the mean square
 dimensions of a nematic Gaussian polymer, and there
 characterise the distribution of chain shapes.  The former is at formation and the latter is that currently pertaining,
that is after any director rotation or changes in $Q$ induced by strain, or due to temperature or
illumination change.  Thus:
 \be  \langle R_{\parallel}R_{\parallel}\rangle = \frac{1}{3} \ell_{\parallel} L, \;\;\;\;
 \langle R_{\perp}R_{\perp}\rangle = \frac{1}{3} \ell_{\perp} L, \;\;\;\; \label{eq:step_def} \ee
 with $\ell_{\parallel}$ and $\ell_{\perp}$ the effective step lengths along and and perpendicular to
 the director, and where $L$ is the chemical arc length of the polymer.
  $\ten{\ell}_0$ and
 $\ten{\ell}_\vec{n}$ depend on the order at formation and that currently pertaining ($Q$).  The
 deformation gradient tensor is $\ten{\lambda}$ and takes the body from its formation state to that
 current.  Since rubber is a soft solid, deformations are strictly at constant volume and thus $\det{\ten{\lambda}\sp{r}} =
 1$.
Such a description of the network chains presumes they are long enough to be Gaussian, albeit
anisotropic.  It is reasonable to make this restriction for otherwise chains would not be dominated
by their random configurations and nor would they be so highly extensible as their experimental
response clearly demands.  With this generic assumption, it then does not matter greatly what kind of
model one adopts for chains -- the Flory philosophy is that local molecular structure is washed out
by randomness and can be encoded by $\ell$ (here two numbers $\ell_{\parallel}$ and $\ell_{\perp}$).
The simplest model  is that of the freely jointed chain with links of length $a$ whereupon
orientational order induces:
 \be
 \frac{\ell_{\parallel}}{a} = (1+2Q), \;
  \frac{\ell_{\perp}}{a} = (1-Q), \; \frac{\det{\ten{\ell}}}{a^3} = (1+2Q)(1-Q)^2 \label{eq:free_joints} \ee
For elastomers over a large range of order parameters and hence also extensions the freely jointed
chain model has proven highly accurate.  It is found \cite{GreveFW} that spontaneous deformations
arising from
 eqn~(\ref{eqn:single}), $\lambda_m = \left( (1+2Q)/(1-Q)\right)^{1/3}$ correlates perfectly with
 independent measurements of $Q$ from optical anisotropy which, along with $\lambda_m$ varies as
 temperature is changed.  Here, distortions $\lambda_m$ are with respect to a high temperature
 reference state where $Q \rightarrow 0$ and $\lambda_m \rightarrow 1$.  Chains can be highly extended by
 high nematic order, attaining for instance a hairpin state which has a much more rapid increase of
 $\ell_{\parallel}$ with $Q$ and hence with $T$ \cite{Sixou}.  In any event the picture of the effective
 step lengths and hence $\det{\ten{\ell}}$ varying with $Q$ persists for all known chain models.  It is unphysical to constrain
 the determinant to a fixed value.  A supplementary argument that has been advanced for the fixing of
 $\det{\ten{\ell}}$ is that this is effectively proportional to the volume in spaced covered by the
 chain and that chains are incompressible, i.e. the $\det{\ten{\ell}}$ cannot change.  The expression
 refers to the extent of the chain, a volume proportional to $N^{3/2}$ since it is a random walk of
 lineal dimension proportional to $n^{1/2}$.  The actual volume occupied by the monomers of the chain
 scales like $N$ (times $a^3$) and thus the density of segments of a particular chain is $\sim
 N^{-1/2}a^{-3}$ -- chains are very dilute in a melt and most of the space they span is occupied by
 other chains.  There is accordingly no volume constraint presented to a given chain.  In fact this
 argument is at the heart of why chains adopt ideal statistics in a melt \cite{deGennes79}.

 What then stops the free energy density (\ref{eqn:single}) minimising at $f = -\infty$ by tending to
 $Q = 1$ where $\det{\ten{\ell}} = 0$ and hence $\ln\left(\det{\ten{\ell}_0}/\det{\ten{\ell}}\right) \rightarrow -\infty$?
One must consider the dominant free energy in the problem, namely that of the underlying nematic
phase rather than the weaker rubber part.  The nematic chains have, in the absence of linking, a
nematic free energy density $f_N(Q)$ that one can show to be additive to the rubber part arising on
linkage.  Model forms arise from simple Landau-de Gennes phenomenology, or from particular molecular
models, for instance the extension of Maier-Saupe nematic theory to worm-like chains.  In fact a
freely jointed rod model of nematic  chains \cite{Abramchuk:87} would have exactly the Maier-Saupe
nematic free energy since the rods are independent, except through their nematic interactions. As
with the chains, it is not vital for our argument what model one adopts.  They all have in common a
free energy of order $k_{\rm B}T_{\rm ni}$ \textit{per monomer} where $T_{\rm ni}$ is a
characteristic temperature where nematic order is lost.  The rubber free energy by contrast is of the
order of $k_{\rm B}T$ per network strand since the prefactor in the free energy density
eqn~(\ref{eqn:single}) is the modulus $\mu = n_{\rm s}k_{\rm B}T$ where $n_{\rm s}$ is the number
density of strands.  Thus the rubber component of the free energy density is of order $1/N$ smaller
than the nematic part.  $N$ is the number of effective step lengths separating crosslink points along
a chain.  $N$ has to be large, $\sim 10 -- 100$ in order that chains are Gaussian and the network is
rubbery, that is highly extensible and dominated by entropy.  The overall dependence of the free
energy on $Q$ is dominated by the minimum determining the stable state of the liquid under the same
conditions.  It is little perturbed by the $Q$-dependent terms arising from the rubber, including the
$\ln\det{\dots}$ term. The rubber adds terms like $\frac{1}{N}(-Q_0^2Q^2 +Q^4)$
\cite{Abramchuk:87,WGV,warnerbook:03} for the ideal case and where spontaneous extension/contraction
has been allowed to occur.  ($Q_0$ is the order at crosslinking.)  These terms cause minor shifts in
the transition temperature and latent heat. In the non-ideal case, involving memory of orientation
pertaining at formation, a term like $-\frac{1}{N}Q$ arises, which resembles the form of an external
field and removes the isotropic high temperature phase $Q$ identically equal to zero altogether.

Strictly speaking, even though the liquid nematic free energy dominates the nematic contributions
from the rubber, the $\ln\det{\dots}$ term would still yield a separate minimum at $Q=1$.  However
the use of the result $\ell_{\perp} = a (1-Q)$ near the point $Q=1$ is also unphysical.  This limit
of perfect orientational order would suggest that $\langle R_{\perp}^2 \rangle \rightarrow 0$.  Only
for chains mathematically narrow and without any directional fluctuations (at $T=0$) could attain
this limit.  In any case the Gaussian approximation of this result will also have long since failed.

How should one proceed?  Physically, one is constrained to being near in $Q$ to a modified nematic
minimum in the free energy.  One can thus reasonably ignore any minimum near $Q=1$ and rely on the
nematic free energy to naturally constrain $Q$-dependence arising in $f_{\rm el}$.  Practically,
unless one is close to the N-I transition where the minimum in $f$ near $Q=0$ is not stiff, one can
take $Q$ to be fixed in magnitude.  If strains are extreme, then one can let $Q$ have the freedom to
respond to perturbations arising from $f_{\rm el}(\ten{\lambda},Q)$ by taking the nematic free energy
to be $\half f''_Q\cdot (Q-Q_{\rm min})^2$ (with $f''_Q$ the curvature of the potential at $Q_{\rm
min}$), which was an approached followed in \cite{GreveFW}.  These authors were concerned changes in
the linear modulus from applied strains inducing changes in $Q$.  This work  also considered the role
of $\ten{\lambda}$ in inducing changes in the biaxial order if the principal stretch direction was
not aligned with $\vec{n}_0$.  Such changes too must be accounted for in the changing
$\ln\det{\dots}$ term that is the subject of this note.

In summary we have argued physically as to why another source of $Q$-dependent free energy terms from
rubber elasticity must not be constrained to take a constant value.  In fact their form relates
closely to well tested models of chain statistics and for the Gaussian theory of nematic rubber
elasticity to hold, these terms must be retained.

\end{document}